\documentclass[10pt,logo,copyright,nonumbering]{nvidiatechreport}
\ifPDFTeX\else\microtypesetup{tracking=false}\fi
\usepackage[numbers]{natbib}

\usepackage{amsmath,amsfonts,bm}

\def\eqref#1{equation~\ref{#1}}

\def\1{\bm{1}}

\DeclareMathAlphabet{\mathsfit}{\encodingdefault}{\sfdefault}{m}{sl}
\SetMathAlphabet{\mathsfit}{bold}{\encodingdefault}{\sfdefault}{bx}{n}

\usepackage{graphicx}
\usepackage{pgfplots}
\usepgfplotslibrary{groupplots}
\pgfplotsset{compat=1.18}
\definecolor{handoffGLM}{HTML}{2864B7}
\definecolor{handoffKimi}{HTML}{D78124}
\definecolor{handoffGPT}{HTML}{8053AC}
\definecolor{handoffQwen}{HTML}{64748B}
\definecolor{frontierGLM}{HTML}{8053AC}
\definecolor{frontierKimi}{HTML}{B91C1C}
\DeclareRobustCommand{\frontierdiamond}{\tikz[baseline=0ex]{\fill[frontierGLM] (0,.11)--(.11,.22)--(.22,.11)--(.11,0)--cycle;}}
\DeclareRobustCommand{\frontiersquare}{\tikz[baseline=0ex]{\fill[frontierKimi] (0,0) rectangle (.19,.19);}}
\DeclareRobustCommand{\handoffline}[2]{\tikz[baseline=-0.5ex]{\draw[color=#1,#2,line width=1pt] (0,0)--(.38,0);}}

\newcounter{handoffpanel}[figure]

\usepackage{booktabs}
\usepackage{multirow}
\usepackage{wrapfig}
\usepackage{array}
\usepackage{amsmath,amssymb}
\usepackage{hyperref}
\usepackage{url}
\usepackage{microtype}

\hypersetup{colorlinks=true,citecolor=blue,linkcolor=blue,urlcolor=blue}
\title{\spilot: Efficient Model Collaboration in Agentic Tasks
}

\author{
Junlong Li\textsuperscript{$\clubsuit$,$\dagger$},
Jindong Jiang\textsuperscript{$\P$},
Yonggan Fu\textsuperscript{$\P$},
Jiarui Yao\textsuperscript{$\spadesuit$,$\dagger$},
Hanrong Ye\textsuperscript{$\P$}\\
\Authfont
Wonmin Byeon,
Peter Belcak,
Hao Zhang,
Pavlo Molchanov\\
\Affilfont
\textsuperscript{$\clubsuit$}The Hong Kong University of Science and Technology\\
\textsuperscript{$\spadesuit$}University of Illinois Urbana-Champaign\\
\textsuperscript{$\P$}Project Leads
}

\renewcommand{\copyrightext}{%
  \footerfont
  \textsuperscript{$\dagger$}Work done during internship at NVIDIA.\\
  \ \ \ \textcopyright\,2026 NVIDIA. All rights reserved.%
}
\newcommand{\spilot}{\textsc{AgentBaton}}

\begin{abstract}

Solving long-horizon agentic tasks requires language agents to interact with their environments over multiple steps, and reliable performance often comes from the expensive \textbf{capable models} available through cloud providers.
Meanwhile, alternative \textbf{efficient models} with inferior performance but a much lower cost are emerging, including lower-tier cloud models and smaller open-weights models deployable on local hardware.
This raises a natural question that remains less explored than traditional query-level routing: can capable and efficient models collaborate within a single long-horizon agentic task to substantially reduce cost while preserving task accuracy? 
In this paper, we systematically investigate key design choices for such collaboration through model handoffs, where one model takes over from another, focusing on handoff direction and context transfer. 
We find that efficient-to-capable handoffs generally underperform capable-only execution
with limited or even no cost reduction, whereas capable-to-efficient handoffs at suitable stages of task progress largely preserve or even improve accuracy while substantially reducing cost.
These findings suggest that a one-time, simple, well-timed handoff strategy is good enough for cost-effective agent collaboration, motivating \textbf{\spilot{}}, a lightweight detector that learns when to hand off from recurring patterns in task progress.
Across 11 model pairs and two representative benchmarks, \spilot{} reduces API costs by 65.9\% on average while largely preserving task accuracy, with accuracy gains of up to 8.8 percentage points over capable-only execution. Experiments across various reasoning-effort settings further show that \spilot{} extends the empirical accuracy--cost Pareto frontier.
\end{abstract}

\begin{document}
\maketitle

\begin{figure}[!ht]
\centering
\includegraphics[width=\linewidth]{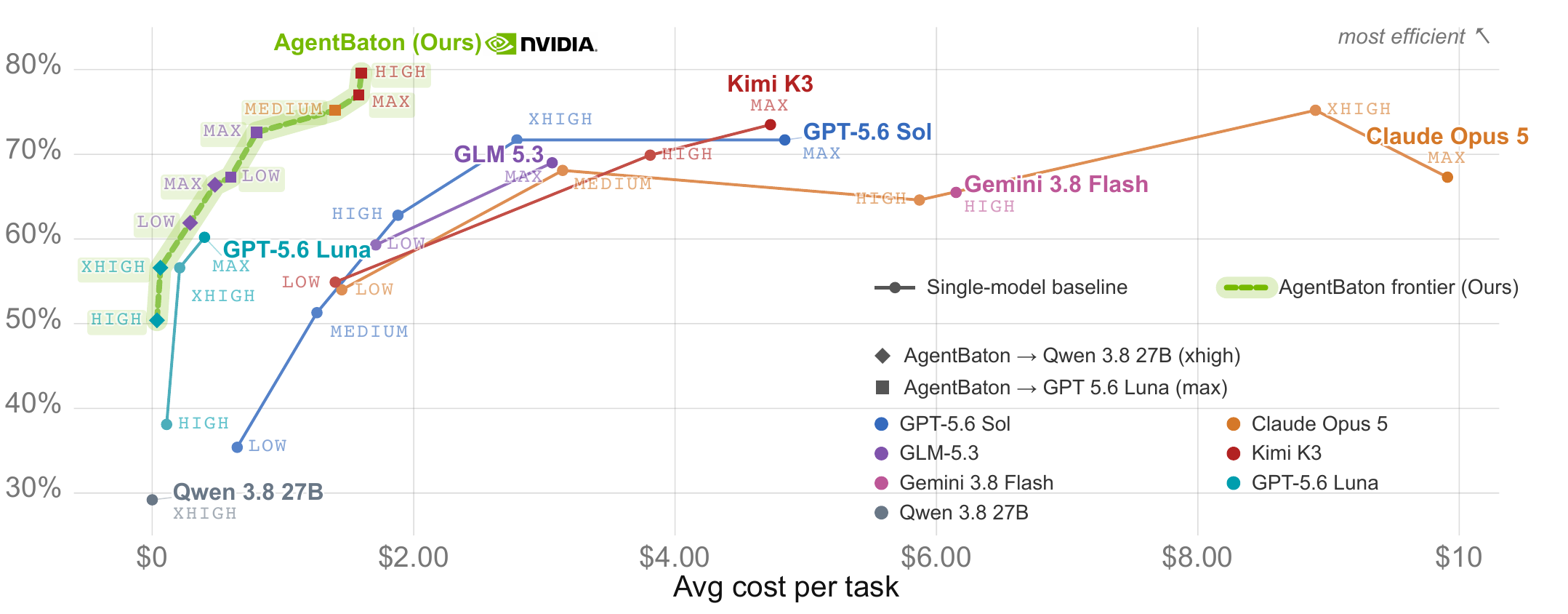}
\caption{Accuracy--cost trade-offs across reasoning-effort settings on DeepSWE v1.1. Solid lines and circles denote single-model baselines. For \spilot{}, diamonds denote handoffs to Qwen 3.8 27B and squares denote handoffs to GPT-5.6 Luna; marker colors identify the source model, as listed in the legend. The green dashed line with a light green band connects the empirical Pareto frontier among evaluated \spilot{} configurations, whose effort labels are highlighted.
For example, the purple \frontierdiamond{} labeled \texttt{MAX} denotes GLM-5.3 at \texttt{max} effort handing off to Qwen 3.8 27B at \texttt{xhigh} effort; the red \frontiersquare{} labeled \texttt{HIGH} denotes Kimi K3 at \texttt{high} effort handing off to GPT-5.6 Luna at \texttt{max} effort.}
\label{fig:reasoning-effort-tradeoff}
\end{figure}

\clearpage

\section{Introduction}
Language-model agents tackle long-horizon tasks through repeated interactions with their environments, using tools, observing feedback, and revising their actions over multiple steps \citep{yao2023react,yang2024sweagent}. Reliable and state-of-the-art performance is often achieved by expensive \textbf{capable models} available through cloud providers \citep{fable51,gpt6astra},
whose repeated invocation can incur substantial inference costs. Meanwhile, alternative \textbf{efficient models}, including lower-tier cloud models \citep{gpt6solluna}
and smaller open-weight models \citep{qwen2026qwen38} deployable on local hardware such as NVIDIA DGX Spark and Apple Mac Studio, offer much lower costs but often achieve lower task accuracy when used alone.


Prior work on combining heterogeneous models has largely focused on query-level routing and cascading for individual requests \citep{chen2023frugalgpt,ong2024routellm,chen2024routerdc,hu2024routerbench}. However, assigning a task to a model solely based on its initial query falls short in long-horizon agentic tasks. In these environments, actions alter the workspace and yield dynamic feedback, meaning the actual capabilities required evolve across turns rather than being determined upfront. This necessitates collaboration throughout the task trajectory. For example, a capable model might draft an initial plan or produce key intermediate results for an efficient model to build upon. Conversely, an efficient model could handle preliminary steps like information gathering before handing off to a capable model for more demanding steps. Both directions are intuitively plausible, leaving open the question of which paradigm best balances accuracy and cost.

Beyond the handoff direction, transferring interaction history introduces another trade-off. Directly including the full history in context ensures the receiving model understand prior progress and reuse existing ideas, but increases the prefill sequence length at handoff and, while that history is retained, the input sequence length in subsequent turns. Moreover, exposing the receiving model to its predecessor's reasoning and actions may bias its subsequent decisions, potentially carrying forward unproductive assumptions or strategies.

To understand how handoff direction and context transfer affect the accuracy--cost trade-off, we evaluate both handoff directions across multiple capable--efficient model pairs on two agentic benchmarks, Terminal Bench 2.1 \citep{merrill2026terminal} and DeepSWE v1.1 \citep{datacurve2026deepswe}. For each direction, we compare two context transfer methods: directly including prior interaction history in the receiving model's context, or providing structured execution history through a log file that the receiving model can inspect as needed using standard commands such as \texttt{cat}, \texttt{grep}, and \texttt{sed}. In both cases, the receiving model continues from the workspace produced by its predecessor.


The experiments reveal an asymmetry between the two handoff directions. Efficient-to-capable handoffs generally underperform capable-only execution, sometimes even at higher cost, despite benefiting from logfile transfer over in-context transfer. In contrast, capable-to-efficient handoffs at suitable stages of task progress can largely preserve capable-only accuracy while substantially reducing cost, although the preferred context-transfer method varies across models. These findings suggest that the progress established by a capable model can enable an efficient model to complete tasks that it struggles to solve independently. This one-way dependency also implies that per-turn routing may be impractical, as frequent switching risks accumulating trajectory errors that the capable model must repeatedly expend effort to recover from.

Our case studies help characterize these handoff opportunities: the capable model has often produced a substantive implementation or artifact and obtained task-relevant feedback, leaving primarily localized repair, validation, or finalization. Motivated by these recurring patterns, we introduce \textbf{\spilot{}}, a lightweight detector trained to identify suitable handoff points from execution history. The capable model begins the task, and \spilot{} evaluates progress after each turn; once it predicts a suitable handoff point, the efficient model takes over and continues until termination. Across 11 model pairs and two benchmarks, \spilot{} reduces API costs by 65.9\% on average while largely preserving task accuracy, with gains of up to 8.8 percentage points over capable-only execution. As shown in Figure~\ref{fig:reasoning-effort-tradeoff}, \spilot{} extends the empirical accuracy--cost Pareto frontier beyond the evaluated single-model configurations across reasoning-effort settings.

To sum up, we make three main contributions:
\begin{itemize}
    \item{We systematically analyze how handoff direction, context transfer, and task progress affect the accuracy--cost trade-offs of model collaboration within long-horizon agentic tasks.}
    \item{We introduce \spilot{}, a lightweight learned detector that identifies suitable capable-to-efficient handoff points online based on recurring patterns in agent execution.}
    \item{We evaluate \spilot{} across multiple model pairs and two agentic benchmarks, demonstrating substantial API cost reductions with largely preserved accuracy and improved accuracy--cost trade-offs across reasoning-effort settings.}
\end{itemize}

\section{Experimental Design}
\label{sec:setup}
\begin{figure}[t]
\centering
\includegraphics[width=\linewidth]{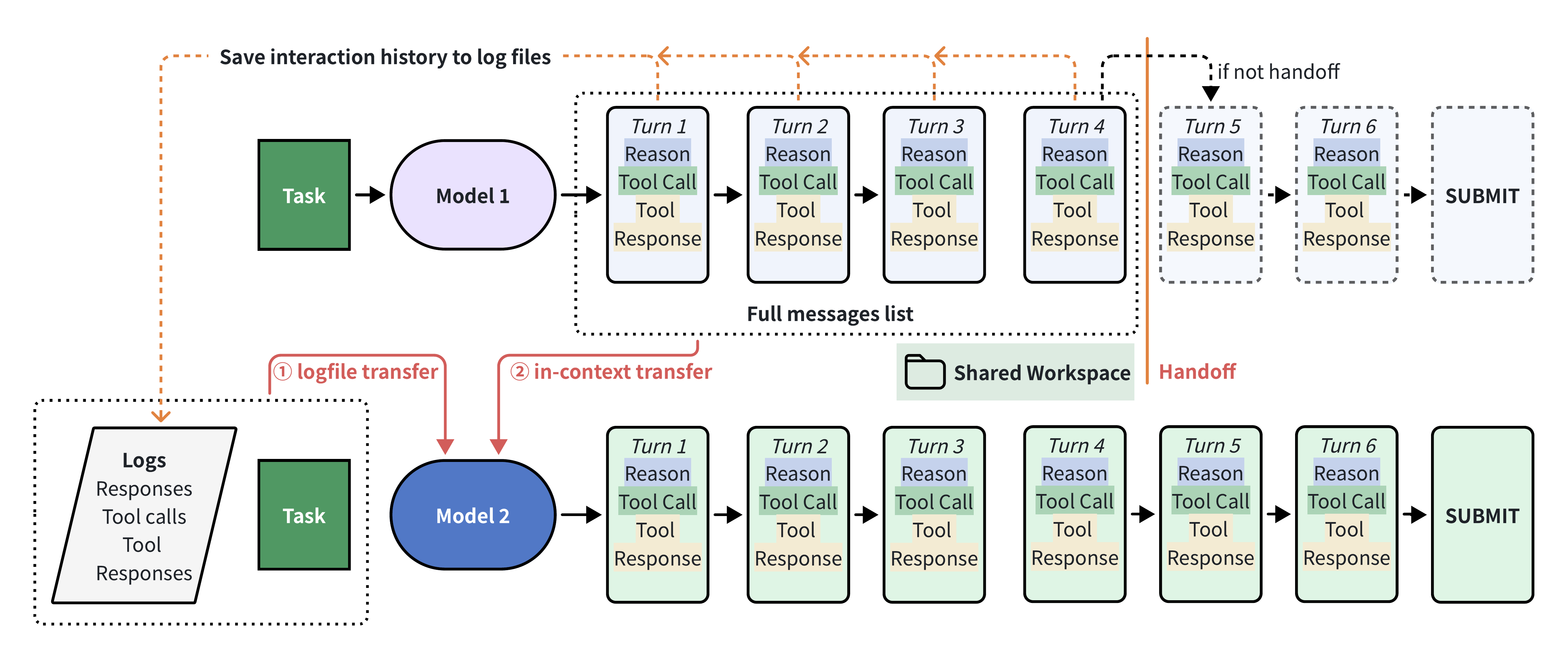}
\caption{Overview of one-time model handoff and context transfer. The receiving model inherits the workspace at handoff and continues the task. Execution history is provided either as an inspectable logfile or directly in context.
}
\label{fig:handoff-overview}
\end{figure}

\paragraph{Model handoff and context transfer.}
For simplicity and controlled analysis, we study model collaboration through one-time handoffs, where an initial model begins a task and transfers control to another after several turns. 
The receiving model then continues until termination, without switching back to the source model. 
The total cost of a handoff run is the sum of the source model's cost before handoff and the receiving model's cost afterward.
The receiving model inherits the workspace state, including all file modifications and artifacts produced by the source model. To transfer the accompanying execution history, we consider two approaches.
1) \textit{In-context transfer} directly includes the pre-handoff conversation in the receiving model's context, preserving message roles, assistant responses, tool calls, and their corresponding tool outputs. When the source model uses a different response format, we convert its visible text and tool interactions into the receiving model's chat format. When the source model does not expose its chain-of-thought, we omit the \texttt{reasoning\_content} field rather than reconstructing unavailable reasoning.
2) \textit{Logfile transfer} instead writes the history to a \texttt{.jsonl} file inside the task environment, where the receiving model can inspect it using its tools. The file contains one JSON record per source-model turn, ordered chronologically. Each record includes the turn index, the original model response, and a list of actions. Each action records the original command and the tool output, including its return code, standard output, and standard error. Under \textit{Logfile transfer}, the receiving model starts a new agent session with the original task instruction, the logfile path, and a textual schema describing the fields and structure of the log. The instruction specifies that records are ordered chronologically and directs the model to inspect the final records first to understand the latest task state. Figure~\ref{fig:handoff-overview} illustrates the handoff process and the two context-transfer approaches.

\paragraph{Benchmarks, Models, Harness and Settings}
We evaluate on Terminal-Bench 2.1 \citep{merrill2026terminal} and DeepSWE v1.1 \citep{datacurve2026deepswe}. Terminal-Bench covers diverse terminal-based workflows, including software engineering, scientific computing, and system administration, whereas DeepSWE focuses on challenging, long-horizon software engineering tasks in active open-source repositories, requiring substantial implementation work evaluated using program-based verifiers. For the analysis experiments, we primarily use GPT-5.6 Sol \citep{openai2026gpt56sol}, Kimi K3 \citep{moonshot2026kimik3}, and GLM-5.3 \citep{zai2026glm53} as capable models, with reasoning effort set to \texttt{max}. We use Qwen3.8-27B \citep{qwen2026qwen38} as the efficient model, with reasoning effort set to \texttt{xhigh}, as it offers strong coding and reasoning capabilities at a scale suitable for deployment on local workstations or PCs. We adopt mini-SWE-agent v2.4.0 \citep{minisweagent}, a minimal yet effective scaffold with a shell-based action interface and a linear interaction history. Its simple agent loop provides a common interface for comparing models and inspecting their trajectories, while capturing the basic interaction pattern of more complex harnesses.
The scaffold is also used in official evaluations of frontier foundation models \citep{deepseekv41flash,claudeopus5,zai2026glm53,moonshot2026kimik3}. For both benchmarks, we impose a two-hour agent execution timeout per task and allow up to 16K newly generated tokens per turn. We use the model providers' recommended sampling parameters where available and otherwise retain the serving defaults.
For cloud models, we calculate cost from the token usage reported for each request and the provider's official token pricing, while assigning zero cost to locally deployable models under our accounting convention. 

\section{Understanding Model Handoffs in Agentic Tasks}
\label{sec:context-direction}
\subsection{Handoff Direction and Context Transfer}
\label{sec:context-transfer}
We first run each of the three capable models and the efficient model independently to obtain single-model baselines and execution trajectories. For a given trajectory, we replay its recorded actions in a task container to restore the workspace state after a selected turn, allowing a receiving model to continue from that checkpoint. We select handoff points at 30\% and 75\% of the source model's completed baseline trajectory to represent relatively early and late stages of execution. At each checkpoint, we evaluate both capable-to-efficient and efficient-to-capable handoffs using the in-context and logfile transfer methods described in Section~\ref{sec:setup}.

The results in Table~\ref{tab:context-transfer} show that the effect of context transfer depends on the handoff direction. For efficient-to-capable handoffs, logfile transfer generally matches or exceeds the accuracy of in-context transfer, while significantly reducing the cost in most cases.
In the reverse direction, in-context transfer is more effective for GLM 5.3 and Kimi K3, achieving higher accuracy in most cases. GPT-5.6 Sol yields a mixed result: logfile transfer has higher accuracy at 30\% on both benchmarks and at 75\% on DeepSWE v1.1, whereas in-context transfer performs better at 75\% on Terminal-Bench. We hypothesize that this difference may partly stem from the availability of reasoning traces: unlike GLM-5.3 and Kimi K3, GPT-5.6 Sol does not expose its full chain-of-thought, which may limit the reasoning context available to the receiving model under in-context transfer.
Thus, we conclude that logfile transfer is more favorable for efficient-to-capable handoffs, while the preferred format for capable-to-efficient handoffs depends on the source model and checkpoint.

\begin{table}[t]
\centering
\caption{Context transfer across handoff directions at 30\% and 75\% of the source model's baseline trajectory. Each cell reports accuracy (\%) / mean recorded API cost per task (USD) in the handoff runs. Bold marks strictly higher accuracy or lower cost between transfer methods at the same checkpoint. GPT, GLM, Kimi, and Qwen denote GPT-5.6 Sol, GLM-5.3, Kimi K3, and Qwen3.8-27B. 
}
\label{tab:context-transfer}
\setlength{\tabcolsep}{3pt}
\resizebox{\linewidth}{!}{%
\begin{tabular}{@{}lrrrrrrrr@{}}
\toprule
& \multicolumn{4}{c}{Terminal-Bench 2.1} & \multicolumn{4}{c}{DeepSWE v1.1}\\
\cmidrule(lr){2-5}\cmidrule(l){6-9}
& \multicolumn{2}{c}{30\%} & \multicolumn{2}{c}{75\%} & \multicolumn{2}{c}{30\%} & \multicolumn{2}{c}{75\%}\\
\cmidrule(lr){2-3}\cmidrule(lr){4-5}\cmidrule(lr){6-7}\cmidrule(l){8-9}
Handoff & In-context & Logfile & In-context & Logfile & In-context & Logfile & In-context & Logfile\\
\midrule
\multicolumn{9}{@{}l}{\textit{Efficient $\rightarrow$ Capable}}\\
Qwen $\to$ GLM & 83.2 / 1.32 & 83.2 / \textbf{1.22} & 68.5 / \textbf{1.19} & \textbf{79.8} / 1.26 & 51.3 / 4.04 & \textbf{62.8} / \textbf{2.93} & 38.9 / 2.61 & \textbf{53.1} / \textbf{1.87}\\
\addlinespace[2pt]
Qwen $\to$ Kimi & 79.8 / 2.91 & \textbf{84.3} / \textbf{1.58} & 70.8 / 3.31 & \textbf{79.8} / \textbf{1.30} & 47.8 / 7.77 & \textbf{62.0} / \textbf{4.45} & 37.2 / 5.76 & \textbf{50.4} / \textbf{2.52}\\
\addlinespace[2pt]
Qwen $\to$ GPT & 83.2 / 1.19 & \textbf{89.9} / \textbf{0.87} & 82.0 / 2.15 & \textbf{85.4} / \textbf{0.85} & \textbf{73.5} / 6.20 & 70.8 / \textbf{5.00} & 58.4 / 8.84 & \textbf{67.3} / \textbf{5.74}\\
\midrule
\multicolumn{9}{@{}l}{\textit{Capable $\rightarrow$ Efficient}}\\
GLM $\to$ Qwen & \textbf{77.5} / 0.15 & 76.4 / 0.15 & \textbf{83.2} / 0.64 & 79.8 / 0.64 & \textbf{61.1} / 0.56 & 54.9 / 0.56 & \textbf{62.8} / 2.07 & 59.3 / 2.07\\
\addlinespace[2pt]
Kimi $\to$ Qwen & \textbf{79.8} / 0.33 & 75.3 / 0.33 & \textbf{87.6} / 1.06 & 82.0 / 1.06 & \textbf{69.9} / 1.11 & 56.6 / 1.11 & 76.1 / 3.34 & 76.1 / 3.34\\
\addlinespace[2pt]
GPT $\to$ Qwen & 64.0 / 0.24 & \textbf{69.7} / 0.24 & \textbf{83.2} / 0.70 & 77.5 / 0.70 & 54.9 / 1.25 & \textbf{56.6} / 1.25 & 66.4 / 3.65 & \textbf{67.3} / 3.65\\
\bottomrule
\end{tabular}%
}
\vspace{-2pt}
\end{table}

\subsection{When Does Handoff Preserve Accuracy?}

\begin{figure}[t]
\centering
\begin{tikzpicture}
\definecolor{glm}{HTML}{2864B7}
\definecolor{kimi}{HTML}{D78124}
\definecolor{gpt}{HTML}{8053AC}
\definecolor{qwen}{HTML}{64748B}
\begin{groupplot}[group style={group size=4 by 1,horizontal sep=.7cm},width=2.55cm,height=3.57cm,scale only axis,xmin=-2,xmax=92,xtick={0,30,60,90},minor xtick={15,45,75},tick label style={font=\fontsize{9}{10}\selectfont},label style={font=\fontsize{9}{10}\selectfont},title style={font=\fontsize{9}{10}\selectfont,yshift=-4pt},axis line style={line width=.6pt},axis lines=left,ymajorgrids,grid style={gray!15},tick align=outside,scaled ticks=false]
\nextgroupplot[ymin=24,ymax=81,title={Accuracy (\%)}]
\addplot[glm,dashed,line width=1.0pt,opacity=.9,forget plot] coordinates {(-2,61.06) (92,61.06)};
\addplot[kimi,dashed,line width=1.0pt,opacity=.9,forget plot] coordinates {(-2,73.45) (92,73.45)};
\addplot[gpt,dashed,line width=1.0pt,opacity=.9,forget plot] coordinates {(-2,68.1) (92,68.1)};
\addplot[qwen,dashed,line width=1.0pt,opacity=.9,forget plot] coordinates {(-2,29.2) (92,29.2)};
\addplot[glm,line width=1.15pt,mark=*,mark size=1.65pt] coordinates {(0,61.06) (15,61.95) (30,62.83) (45,56.64) (60,49.56) (75,53.1) (90,52.21)};
\addplot[kimi,line width=1.15pt,mark=*,mark size=1.65pt] coordinates {(0,73.45) (15,64.6) (30,61.95) (45,54.87) (60,52.21) (75,50.44) (90,51.33)};
\addplot[gpt,line width=1.15pt,mark=*,mark size=1.65pt] coordinates {(0,68.1) (15,68.14) (30,70.8) (45,67.26) (60,67.26) (75,67.26) (90,66.37)};
\nextgroupplot[ymin=-0.18,ymax=6.4,title={Cost (USD)},yticklabel style={/pgf/number format/fixed,/pgf/number format/precision=1}]
\addplot[glm,dashed,line width=1.0pt,opacity=.9,forget plot] coordinates {(-2,3.162) (92,3.162)};
\addplot[kimi,dashed,line width=1.0pt,opacity=.9,forget plot] coordinates {(-2,4.726) (92,4.726)};
\addplot[gpt,dashed,line width=1.0pt,opacity=.9,forget plot] coordinates {(-2,5.32) (92,5.32)};
\addplot[qwen,dashed,line width=1.0pt,opacity=.9,forget plot] coordinates {(-2,0.0) (92,0.0)};
\addplot[glm,line width=1.15pt,mark=*,mark size=1.65pt] coordinates {(0,3.162) (15,3.062) (30,2.932) (45,2.644) (60,2.211) (75,1.867) (90,2.029)};
\addplot[kimi,line width=1.15pt,mark=*,mark size=1.65pt] coordinates {(0,4.726) (15,4.774) (30,4.451) (45,3.923) (60,3.175) (75,2.519) (90,2.49)};
\addplot[gpt,line width=1.15pt,mark=*,mark size=1.65pt] coordinates {(0,5.32) (15,5.024) (30,4.995) (45,5.229) (60,5.388) (75,5.744) (90,5.29)};
\nextgroupplot[ymin=24,ymax=81,title={Accuracy (\%)}]
\addplot[glm,dashed,line width=1.0pt,opacity=.9,forget plot] coordinates {(-2,61.06) (92,61.06)};
\addplot[kimi,dashed,line width=1.0pt,opacity=.9,forget plot] coordinates {(-2,73.45) (92,73.45)};
\addplot[gpt,dashed,line width=1.0pt,opacity=.9,forget plot] coordinates {(-2,68.1) (92,68.1)};
\addplot[qwen,dashed,line width=1.0pt,opacity=.9,forget plot] coordinates {(-2,29.2) (92,29.2)};
\addplot[glm,line width=1.15pt,mark=*,mark size=1.65pt] coordinates {(0,29.2) (15,61.06) (30,61.1) (45,61.95) (60,61.95) (75,62.83) (90,62.83)};
\addplot[kimi,line width=1.15pt,mark=*,mark size=1.65pt] coordinates {(0,29.2) (15,61.95) (30,69.91) (45,72.57) (60,76.11) (75,76.11) (90,76.99)};
\addplot[gpt,line width=1.15pt,mark=*,mark size=1.65pt] coordinates {(0,29.2) (15,47.79) (30,56.64) (45,61.95) (60,67.26) (75,67.26) (90,68.14)};
\nextgroupplot[ymin=-0.18,ymax=6.4,title={Cost (USD)},yticklabel style={/pgf/number format/fixed,/pgf/number format/precision=1}]
\addplot[glm,dashed,line width=1.0pt,opacity=.9,forget plot] coordinates {(-2,3.162) (92,3.162)};
\addplot[kimi,dashed,line width=1.0pt,opacity=.9,forget plot] coordinates {(-2,4.726) (92,4.726)};
\addplot[gpt,dashed,line width=1.0pt,opacity=.9,forget plot] coordinates {(-2,5.32) (92,5.32)};
\addplot[qwen,dashed,line width=1.0pt,opacity=.9,forget plot] coordinates {(-2,0.0) (92,0.0)};
\addplot[glm,line width=1.15pt,mark=*,mark size=1.65pt] coordinates {(0,0.0) (15,0.196) (30,0.557) (45,0.988) (60,1.495) (75,2.072) (90,2.719)};
\addplot[kimi,line width=1.15pt,mark=*,mark size=1.65pt] coordinates {(0,0.0) (15,0.36) (30,1.105) (45,1.82) (60,2.56) (75,3.343) (90,4.19)};
\addplot[gpt,line width=1.15pt,mark=*,mark size=1.65pt] coordinates {(0,0.0) (15,0.4549) (30,1.25) (45,1.996) (60,2.7713) (75,3.65) (90,4.6914)};
\end{groupplot}
\node[font=\fontsize{10}{11}\selectfont] at ([yshift=-.68cm]group c1r1.south) {(a)};
\node[font=\fontsize{10}{11}\selectfont] at ([yshift=-.68cm]group c2r1.south) {(b)};
\node[font=\fontsize{10}{11}\selectfont] at ([yshift=-.68cm]group c3r1.south) {(c)};
\node[font=\fontsize{10}{11}\selectfont] at ([yshift=-.68cm]group c4r1.south) {(d)};
\path (group c1r1.north west) -- (group c2r1.north east) coordinate[midway] (row1);
\node[font=\fontsize{10}{11}\selectfont\bfseries] at ([yshift=.7cm]row1) {Efficient $\to$ Capable};
\path (group c1r1.south west) -- (group c2r1.south east) coordinate[midway] (xlabel1);
\node[font=\fontsize{10}{11}\selectfont] at ([yshift=-1.05cm]xlabel1) {Handoff progress (\%)};
\path (group c3r1.north west) -- (group c4r1.north east) coordinate[midway] (row2);
\node[font=\fontsize{10}{11}\selectfont\bfseries] at ([yshift=.7cm]row2) {Capable $\to$ Efficient};
\path (group c3r1.south west) -- (group c4r1.south east) coordinate[midway] (xlabel2);
\node[font=\fontsize{10}{11}\selectfont] at ([yshift=-1.05cm]xlabel2) {Handoff progress (\%)};
\end{tikzpicture}
\caption[Handoff progress on DeepSWE v1.1]{Accuracy and mean API cost across handoff progress on DeepSWE v1.1. Colors identify the capable model: \handoffline{handoffGLM}{solid} GLM-5.3, \handoffline{handoffKimi}{solid} Kimi K3, and \handoffline{handoffGPT}{solid} GPT-5.6 Sol; the efficient model is Qwen 3.8 27B. Solid curves show handoffs using logfile transfer in (a--b); in (c--d), GLM-5.3 and Kimi K3 use in-context transfer, while GPT-5.6 Sol uses logfile transfer. Horizontal dashed lines (\handoffline{handoffGLM}{dashed}\,\handoffline{handoffKimi}{dashed}\,\handoffline{handoffGPT}{dashed}) show the corresponding capable-only baselines; \handoffline{handoffQwen}{dashed} denotes Qwen-only execution. Progress is relative to the source model's completed baseline trajectory; 0\% denotes receiving-model-only execution.
}
\label{fig:handoff-progress}
\setcounter{handoffpanel}{0}
\refstepcounter{handoffpanel}\label{fig:handoff-e2c-deepswe-accuracy}
\refstepcounter{handoffpanel}\label{fig:handoff-e2c-deepswe-cost}
\refstepcounter{handoffpanel}\label{fig:handoff-c2e-deepswe-accuracy}
\refstepcounter{handoffpanel}\label{fig:handoff-c2e-deepswe-cost}
\vspace{-2pt}
\end{figure}
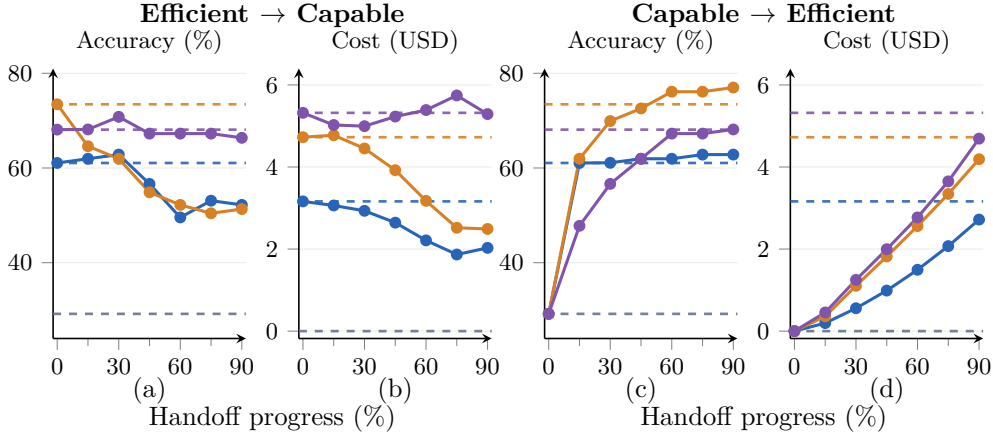

We next ask at what stages of task progress a handoff can preserve capable-only accuracy while substantially reducing cost. Building on the context-transfer comparison, we evaluate both handoff directions at 15\%, 30\%, 45\%, 60\%, 75\%, and 90\% of the source model's completed baseline trajectory. We use logfile transfer for efficient-to-capable handoffs and in-context transfer for GLM-5.3 and Kimi K3 handing off to Qwen 3.8 27B. For GPT-5.6 Sol handing off to Qwen 3.8 27B, we use logfile transfer, which achieves higher accuracy in most comparisons in Section~\ref{sec:context-transfer}. Figure~\ref{fig:handoff-progress} reports DeepSWE v1.1 accuracy and mean cost per task. The 0\% endpoint corresponds to running the receiving model alone. Terminal-Bench 2.1 results are reported in Appendix~\ref{sec:terminalbench-handoff}.

\paragraph{Efficient-to-capable handoffs.}
Figures~\ref{fig:handoff-e2c-deepswe-accuracy} and~\ref{fig:handoff-e2c-deepswe-cost} show that starting with the efficient model does not consistently preserve capable-only accuracy while substantially reducing cost. On DeepSWE v1.1, later handoffs to GLM-5.3 and Kimi K3 generally reduce cost but also lose accuracy. Handoffs to GPT-5.6 Sol can retain comparable accuracy, yet offer limited savings: at 30\% progress, accuracy reaches 70.80\%, compared with 68.10\% for GPT-5.6 Sol alone, while cost falls by only 6.1\%. Thus, delaying the capable model's involvement does not provide a consistent way to achieve substantial savings without sacrificing accuracy. These results also imply that fine-grained or per-turn routing may face steep hurdles, as frequent switching can amplify coordination and recovery overheads.

\paragraph{Capable-to-efficient handoffs.}
Figures~\ref{fig:handoff-c2e-deepswe-accuracy} and~\ref{fig:handoff-c2e-deepswe-cost} show more favorable trade-offs in the reverse direction: at suitable stages of progress, capable-to-efficient handoffs can approach capable-only accuracy while substantially reducing cost. With Kimi K3 as the source model, handing off at 45\% reaches 72.57\% accuracy, close to the capable-only baseline of 73.45\%, while reducing cost by 61.5\%. With GPT-5.6 Sol, logfile transfer at 60\% reaches 67.26\% accuracy versus 68.10\% for this capable model alone, with 47.9\% lower cost. GLM-5.3 can hand off at 30\% and attain 61.1\% accuracy, matching its capable-only baseline, while reducing cost by 82.4\%. These results suggest that progress made by a capable model can enable an efficient model to complete the remaining work.

\subsection{Patterns of Successful Handoffs}
\label{sec:case_study}
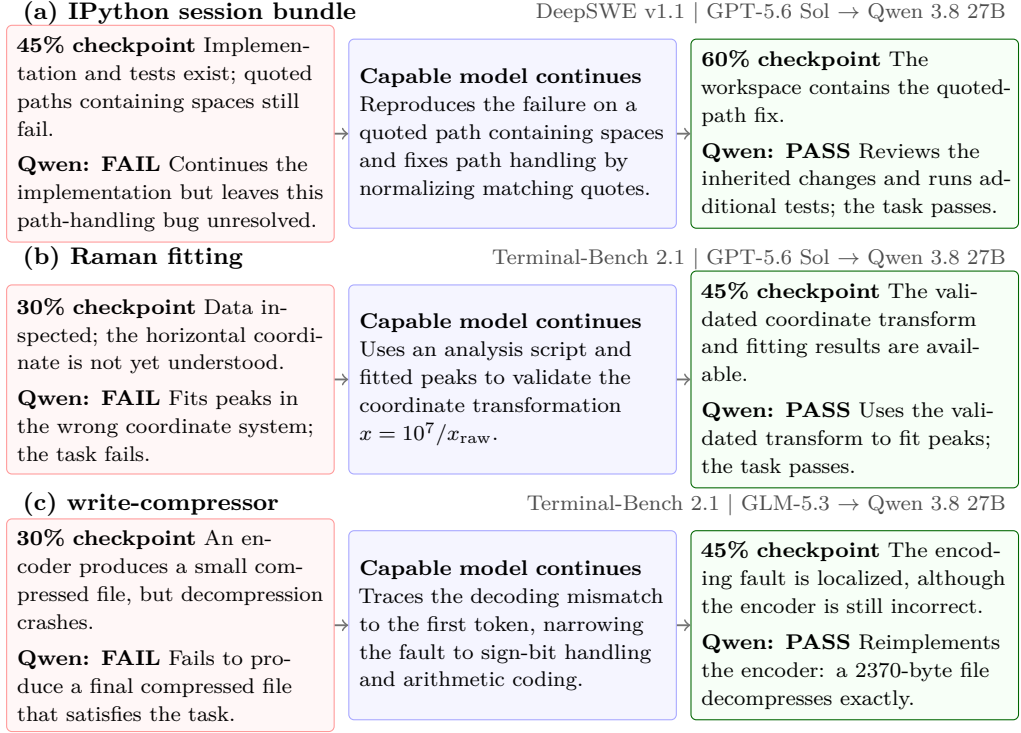
\begin{figure}[t]
\centering
\begin{tikzpicture}[
 font=\fontsize{8}{10.5}\selectfont,
 state/.style={draw=black!25,fill=black!2,rounded corners=2pt,text width=4.05cm,minimum height=2.5cm,align=left,inner sep=4pt},
 progress/.style={state,draw=blue!35,fill=blue!4},
 fail/.style={state,draw=red!40,fill=red!3,minimum height=2.5cm},
 pass/.style={state,draw=green!40!black,fill=green!3,minimum height=2.5cm},
 flow/.style={->,line width=.7pt,draw=black!55}]
\begin{scope}[yshift=0.0cm]
\node[anchor=west,font=\fontsize{9}{11}\selectfont\bfseries] at (-.05,1.60) {(a) IPython session bundle};
\node[anchor=east,font=\fontsize{8}{9}\selectfont,text=black!70] at (13.2,1.60) {DeepSWE v1.1 $\mid$ GPT-5.6 Sol $\to$ Qwen 3.8 27B};
\node[fail] (early) at (2.025,0) {\textbf{45\% checkpoint} Implementation and tests exist; quoted paths containing spaces still fail.\\[3pt] \textbf{Qwen: FAIL} Continues the implementation but leaves this path-handling bug unresolved.};
\node[progress] (change) at (6.55,0) {\textbf{Capable model continues}\\ Reproduces the failure on a quoted path containing spaces and fixes path handling by normalizing matching quotes.};
\node[pass] (late) at (11.075,0) {\textbf{60\% checkpoint} The workspace contains the quoted-path fix.\\[3pt] \textbf{Qwen: PASS} Reviews the inherited changes and runs additional tests; the task passes.};
\draw[flow] (early.east)--(change.west);
\draw[flow] (change.east)--(late.west);
\end{scope}
\begin{scope}[yshift=-3.25cm]
\node[anchor=west,font=\fontsize{9}{11}\selectfont\bfseries] at (-.05,1.60) {(b) Raman fitting};
\node[anchor=east,font=\fontsize{8}{9}\selectfont,text=black!70] at (13.2,1.60) {Terminal-Bench 2.1 $\mid$ GPT-5.6 Sol $\to$ Qwen 3.8 27B};
\node[fail] (early) at (2.025,0) {\textbf{30\% checkpoint} Data inspected; the horizontal coordinate is not yet understood.\\[3pt] \textbf{Qwen: FAIL} Fits peaks in the wrong coordinate system; the task fails.};
\node[progress] (change) at (6.55,0) {\textbf{Capable model continues}\\ Uses an analysis script and fitted peaks to validate the coordinate transformation $x=10^7/x_{\mathrm{raw}}$.};
\node[pass] (late) at (11.075,0) {\textbf{45\% checkpoint} The validated coordinate transform and fitting results are available.\\[3pt] \textbf{Qwen: PASS} Uses the validated transform to fit peaks; the task passes.};
\draw[flow] (early.east)--(change.west);
\draw[flow] (change.east)--(late.west);
\end{scope}
\begin{scope}[yshift=-6.50cm]
\node[anchor=west,font=\fontsize{9}{11}\selectfont\bfseries] at (-.05,1.60) {(c) write-compressor};
\node[anchor=east,font=\fontsize{8}{9}\selectfont,text=black!70] at (13.2,1.60) {Terminal-Bench 2.1 $\mid$ GLM-5.3 $\to$ Qwen 3.8 27B};
\node[fail] (early) at (2.025,0) {\textbf{30\% checkpoint} An encoder produces a small compressed file, but decompression crashes.\\[3pt] \textbf{Qwen: FAIL} Fails to produce a final compressed file that satisfies the task.};
\node[progress] (change) at (6.55,0) {\textbf{Capable model continues}\\ Traces the decoding mismatch to the first token, narrowing the fault to sign-bit handling and arithmetic coding.};
\node[pass] (late) at (11.075,0) {\textbf{45\% checkpoint} The encoding fault is localized, although the encoder is still incorrect.\\[3pt] \textbf{Qwen: PASS} Reimplements the encoder: a 2370-byte file decompresses exactly.};
\draw[flow] (early.east)--(change.west);
\draw[flow] (change.east)--(late.west);
\end{scope}
\end{tikzpicture}
\caption{Unsuccessful and successful capable-to-efficient handoffs. Each row shows an earlier checkpoint, the capable model's intervening progress, and the next checkpoint. Left and right boxes report separate Qwen 3.8 27B continuations. Percentages refer to the capable-only trajectory.}
\label{fig:handoff-cases}
\vspace{-2pt}
\end{figure}

The checkpoint experiments show that handoff timing matters, but do not explain what task progress enables an efficient model to take over successfully. To investigate this question, we analyze trajectories where Qwen 3.8 27B fails after an earlier handoff but succeeds after the next one, comparing the capable model's intervening actions and tool feedback to identify recurring patterns. Figure~\ref{fig:handoff-cases} presents three illustrative cases across two benchmarks and two capable models, all of which fail under Qwen-only execution and succeed under capable-only execution. The key changes include (a) fixing a quoted-path handling issue, (b) validating the coordinate transformation, and (c) localizing an encoding fault that helps Qwen reimplement a still-incorrect encoder.

The analysis suggests three recurring signs of handoff readiness: 1) a substantive implementation, runnable prototype, or task-required artifact is available; 2) actual execution, testing, or inspection has provided task-relevant feedback; and 3) the remaining work has become more focused, often involving localized diagnosis, repair, validation, or final-output preparation. Crucially, this feedback need not be 100\% positive; even a failed execution is informative if it narrows down the unresolved problem, as demonstrated in the compressor case. 
These observations suggest that readiness depends on both the progress already established and the nature of the remaining work, rather than elapsed turns alone.

\section{\spilot{}}
\label{sec:spilot}

\begin{wraptable}{r}{0.53\textwidth}
\vspace{-\intextsep}
\centering
\caption{Prompt-based handoff to Qwen 3.8 27B on DeepSWE v1.1. Handoffs reports the number of tasks with a recorded handoff; cost is mean API cost/task.}
\vspace{8pt}
\label{tab:prompt-handoff}
\small
\setlength{\tabcolsep}{10pt}
\begin{tabular}{@{}lrrr@{}}
\toprule
Capable model & Handoffs & Acc. (\%) & Cost (\$)\\
\midrule
GLM-5.3 & 6/113 & 54.9 & 2.54\\
Kimi K3 & 0/113 & 64.6 & 4.89\\
GPT-5.6 Sol & 113/113 & 62.8 & 3.75\\
\bottomrule
\end{tabular}
\vspace{-16pt}
\end{wraptable}

The patterns identified in Section~\ref{sec:case_study} motivate a natural question: can we train a lightweight model to recognize suitable handoff points during a capable model's execution? Such a detector would enable the system to transfer control to an efficient model when the observed progress indicates that it is ready to take over.

\subsection{Prompt-based handoff}

Before training a detector, we test whether the capable model can decide when to hand off through prompting alone. We prompt the capable model to transfer control when it recognizes the handoff-readiness patterns identified in Section~\ref{sec:case_study}. Table~\ref{tab:prompt-handoff} shows that this behavior varies substantially across models: Kimi K3 never hands off, while GLM-5.3 does so on only 6 of 113 tasks. Thus, a handoff instruction does not reliably induce models to relinquish control, limiting opportunities to reduce capable-model usage. GPT-5.6 Sol does hand off on every task, but its mean cost remains \$3.75, compared with \$5.32 for capable-only execution. Even consistent handoff behavior therefore does not ensure a favorable accuracy--cost trade-off. These results motivate a dedicated detector that explicitly evaluates handoff readiness, rather than relying on the executing model to decide when to transfer control.

\subsection{Training \spilot{}}
\paragraph{Data collection and annotation.}
We collect agent trajectories on a randomly sampled subset of 2,000 tasks from SWE-rebench V2 \citep{badertdinov2026swerebench}. For each task, we randomly select two distinct capable models from a pool of five: GPT-6 Astra \citep{openai2026gpt6astra}, GPT-5.6 Sol \citep{openai2026gpt56sol}, Claude Opus 5 \citep{claudeopus5}, Kimi K3 \citep{moonshot2026kimik3}, and GLM-5.3 \citep{zai2026glm53}. We use \texttt{medium} reasoning effort for GPT-6 Astra and \texttt{max} for the other models. Each selected model independently attempts the task using mini-swe-agent, yielding 4,000 trajectories. We use Codex\footnote{\url{https://openai.com/codex/}} to annotate suitable handoff points in these trajectories. The annotation criteria generally follow the qualitative patterns identified in Section~\ref{sec:case_study}: a substantive implementation or task-required artifact exists, task-relevant execution or inspection provides concrete feedback, and the remaining work is primarily localized diagnosis, repair, or validation. Codex reviews each trajectory retrospectively to better locate the earliest suitable handoff point. Trajectories without a clear transition receive a separate no-clear-handoff label. Appendix~\ref{sec:handoff-annotation} provides the annotation protocol, including evidence restrictions and turn indexing.

\paragraph{Training details.}
We initialize \spilot{} from Qwen 3.5 2B \citep{qwen2026qwen35_2b} for low-overhead detection at each turn.\footnote{We also train a Qwen 3.5 9B \citep{qwen2026qwen35_9b} variant but observe no clear improvement.} The input is a multi-turn conversation consisting of a system instruction that guides pattern detection, the original task, and the capable model's actions and tool observations, interleaved with previous handoff decisions. For a trajectory whose annotated handoff occurs at turn $h$, we supervise every decision up to and including turn $h$: the model is trained to output the JSON object \texttt{\{"judgement":\allowbreak"HANDOFF\_NOT\_READY"\}} for turns before $h$ and \texttt{\{"judgement":\allowbreak"HANDOFF\_READY"\}} at turn $h$.
We also train on all subsequent turns. Each such turn yields an additional example in which the target response for the current turn is set to \texttt{\{"judgement":\allowbreak"HANDOFF\_READY"\}}. All preceding decision responses are set to \texttt{\{"judgement":\allowbreak"HANDOFF\_NOT\_READY"\}} and excluded from the loss calculation.
These additional positive examples help balance the decision labels and simulate missed handoff opportunities, encouraging the detector to trigger a handoff at a later turn even if it previously failed to do so.
We train for two epochs with a maximum sequence length of 32K tokens. We split and truncate overlong sequences and pack shorter ones together to reduce padding and improve computational efficiency. We train with NeMo Automodel \citep{nemo-automodel} on 16 NVIDIA H100 GPUs using FSDP2, bfloat16 precision, and activation checkpointing. The global batch size is 16 packed sequences, with one sequence per GPU. We use AdamW with a learning rate of $10^{-5}$, $\beta_1=0.9$, $\beta_2=0.95$, $\epsilon=10^{-8}$, and weight decay $0.1$.




\paragraph{Online Handoff Detection}
\begin{wrapfigure}{r}{0.50\textwidth}
\vspace{-\intextsep}
\centering
\includegraphics[width=\linewidth]{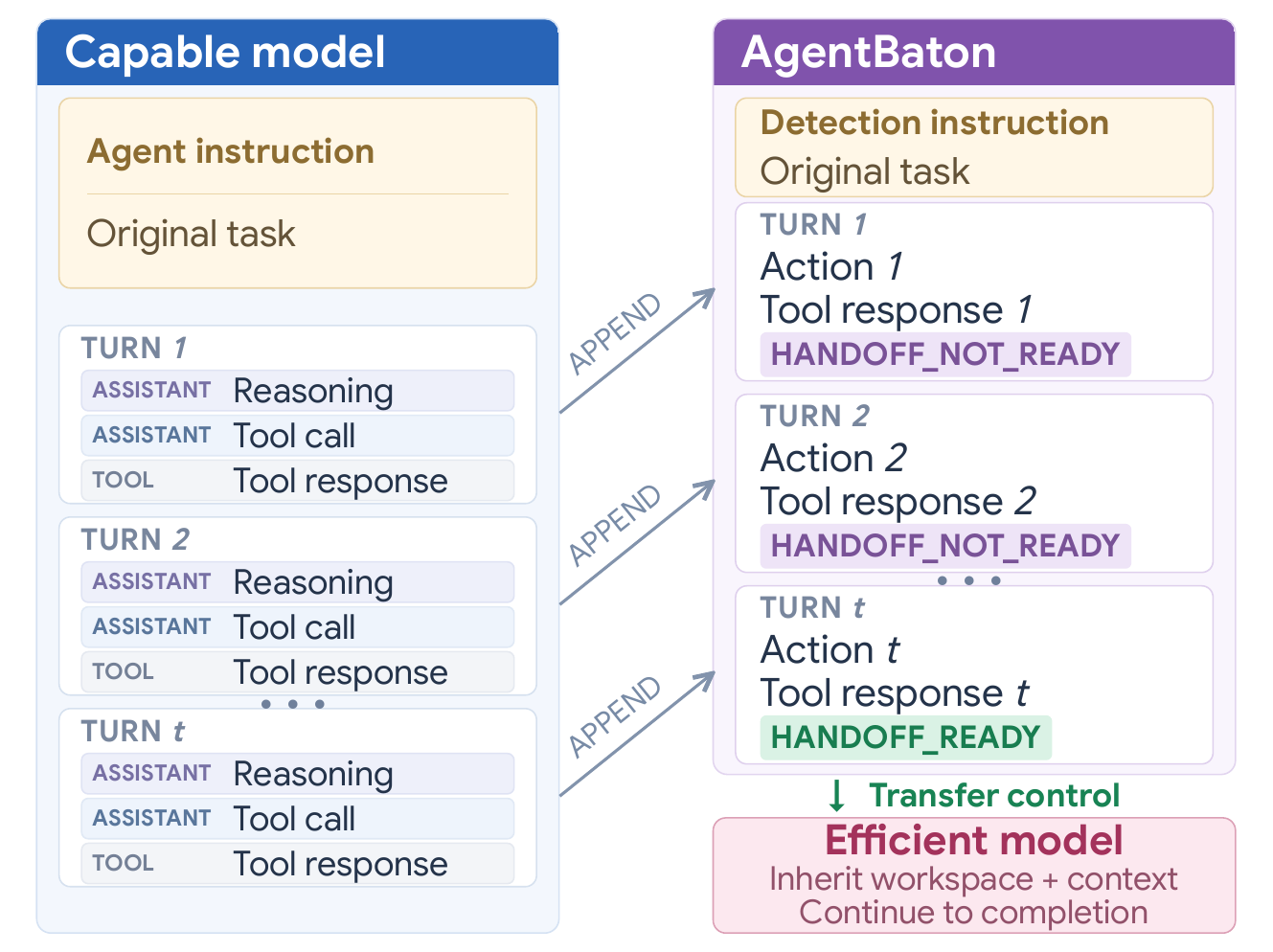}
\caption{Online detection in \spilot{} with separate, growing message histories.
}
\label{fig:online-handoff}
\end{wrapfigure}
At inference time, the capable model begins the task. After each turn, \spilot{} processes the updated trajectory and predicts whether a suitable handoff point has been reached.
When the context exceeds this limit, we discard the oldest half of the historical turns while always retaining the system instruction and original task. A \texttt{HANDOFF\_NOT\_READY} decision leaves control with the capable model; the first \texttt{HANDOFF\_READY} decision transfers control to the efficient model, which inherits the current workspace and continues until termination using the context-transfer method selected for that model pair. Figure~\ref{fig:online-handoff} illustrates this process. Across 360 evaluations on real inputs ranging from 2K to 121K tokens, \spilot{} achieves a median latency of 338\,ms and a 95th-percentile latency of 1.11\,s on a single NVIDIA H100 GPU, demonstrating low latency.

\subsection{Experimental Results}
\begin{table}[t]
\centering
\caption{\spilot{} results. We use logfile transfer when the source model is GPT-5.6 Sol or GPT-5.6 Luna, and in-context transfer for all other source models. Subscripts give relative changes from capable-only: green is better, red worse. Handoff is the mean detected progress (\%), shared across receiving models for a given source. Dashes denote single-model runs. 
}
\label{tab:spilot-results}
\footnotesize
\setlength{\tabcolsep}{2pt}
\setlength{\aboverulesep}{1pt}
\setlength{\belowrulesep}{1pt}
\setlength{\abovecaptionskip}{3pt}
\setlength{\belowcaptionskip}{3pt}
\renewcommand{\arraystretch}{0.92}
\begin{tabular}{@{}llrrrrrr@{}}
\toprule
 & & \multicolumn{3}{c}{DeepSWE v1.1} & \multicolumn{3}{c}{Terminal-Bench 2.1} \\
\cmidrule(lr){3-5}\cmidrule(l){6-8}
Capable model & Efficient model & \shortstack{Handoff\\(\%)} & Acc. (\%) $\uparrow$ & Cost (\$) $\downarrow$ & \shortstack{Handoff\\(\%)} & Acc. (\%) $\uparrow$ & Cost (\$) $\downarrow$ \\
\midrule
--- & Qwen 3.8 27B & --- & 29.2 & 0.00 & --- & 69.7 & 0.00 \\
--- & GPT-5.6 Luna & --- & 60.2 & 0.40 & --- & 82.2 & 0.07 \\
\midrule
GPT-5.6 Sol & --- & --- & 68.1 & 5.32 & --- & 87.6 & 0.97 \\
 & Qwen 3.8 27B & \multirow{2}{*}{26.6} & 60.2\textsubscript{\textcolor{red!75!black}{-11.6\%}} & 1.02\textsubscript{\textcolor{green!45!black}{-80.8\%}} & \multirow{2}{*}{58.4} & 82.0\textsubscript{\textcolor{red!75!black}{-6.4\%}} & 0.43\textsubscript{\textcolor{green!45!black}{-55.7\%}} \\
 & GPT-5.6 Luna &  & 67.3\textsubscript{\textcolor{red!75!black}{-1.2\%}} & 1.39\textsubscript{\textcolor{green!45!black}{-73.9\%}} &  & 82.0\textsubscript{\textcolor{red!75!black}{-6.4\%}} & 0.47\textsubscript{\textcolor{green!45!black}{-51.5\%}} \\
\midrule
Claude Opus 5 & --- & --- & 67.3 & 9.91 & --- & 84.3 & 2.83 \\
 & Qwen 3.8 27B & \multirow{2}{*}{36.6} & 61.9\textsubscript{\textcolor{red!75!black}{-8.0\%}} & 2.61\textsubscript{\textcolor{green!45!black}{-73.7\%}} & \multirow{2}{*}{55.6} & 87.6\textsubscript{\textcolor{green!45!black}{+3.9\%}} & 1.13\textsubscript{\textcolor{green!45!black}{-60.1\%}} \\
 & GPT-5.6 Luna &  & 76.1\textsubscript{\textcolor{green!45!black}{+13.1\%}} & 2.92\textsubscript{\textcolor{green!45!black}{-70.5\%}} &  & 86.5\textsubscript{\textcolor{green!45!black}{+2.6\%}} & 1.17\textsubscript{\textcolor{green!45!black}{-58.7\%}} \\
\midrule
GLM-5.3 & --- & --- & 61.1 & 3.16 & --- & 86.5 & 1.02 \\
 & Qwen 3.8 27B & \multirow{2}{*}{28.6} & 58.4\textsubscript{\textcolor{red!75!black}{-4.4\%}} & 0.49\textsubscript{\textcolor{green!45!black}{-84.5\%}} & \multirow{2}{*}{55.7} & 80.9\textsubscript{\textcolor{red!75!black}{-6.5\%}} & 0.27\textsubscript{\textcolor{green!45!black}{-73.5\%}} \\
 & GPT-5.6 Luna &  & 66.4\textsubscript{\textcolor{green!45!black}{+8.7\%}} & 0.83\textsubscript{\textcolor{green!45!black}{-73.7\%}} &  & 79.8\textsubscript{\textcolor{red!75!black}{-7.7\%}} & 0.33\textsubscript{\textcolor{green!45!black}{-67.6\%}} \\
\midrule
Kimi K3 & --- & --- & 73.5 & 4.73 & --- & 86.5 & 1.50 \\
 & Qwen 3.8 27B & \multirow{2}{*}{35.3} & 72.6\textsubscript{\textcolor{red!75!black}{-1.2\%}} & 1.28\textsubscript{\textcolor{green!45!black}{-72.9\%}} & \multirow{2}{*}{64.0} & 86.5\textsubscript{\textcolor{black!55}{0.0\%}} & 0.57\textsubscript{\textcolor{green!45!black}{-62.0\%}} \\
 & GPT-5.6 Luna &  & 77.0\textsubscript{\textcolor{green!45!black}{+4.8\%}} & 1.58\textsubscript{\textcolor{green!45!black}{-66.6\%}} &  & 80.9\textsubscript{\textcolor{red!75!black}{-6.5\%}} & 0.62\textsubscript{\textcolor{green!45!black}{-58.7\%}} \\
\midrule
Gemini 3.8 Flash & --- & --- & 65.5 & 6.15 & --- & 86.5 & 1.02 \\
 & Qwen 3.8 27B & \multirow{2}{*}{41.1} & 61.9\textsubscript{\textcolor{red!75!black}{-5.5\%}} & 1.21\textsubscript{\textcolor{green!45!black}{-80.3\%}} & \multirow{2}{*}{68.9} & 86.5\textsubscript{\textcolor{black!55}{0.0\%}} & 0.51\textsubscript{\textcolor{green!45!black}{-50.0\%}} \\
 & GPT-5.6 Luna &  & 73.5\textsubscript{\textcolor{green!45!black}{+12.2\%}} & 1.71\textsubscript{\textcolor{green!45!black}{-72.2\%}} &  & 85.4\textsubscript{\textcolor{red!75!black}{-1.3\%}} & 0.56\textsubscript{\textcolor{green!45!black}{-45.1\%}} \\
\midrule
GPT-5.6 Luna & Qwen 3.8 27B & 32.6 & 55.8\textsubscript{\textcolor{red!75!black}{-7.3\%}} & 0.10\textsubscript{\textcolor{green!45!black}{-75.0\%}} & 64.4 & 77.5\textsubscript{\textcolor{red!75!black}{-5.7\%}} & 0.04\textsubscript{\textcolor{green!45!black}{-42.9\%}} \\
\bottomrule
\end{tabular}
\vspace{-11pt}
\end{table}


We further extend the model pool in Section~\ref{sec:context-direction} with two additional capable models, Claude Opus 5 \citep{claudeopus5} and Gemini 3.8 Flash \citep{gdm2026gemini3_8}, and another efficient model, GPT-5.6 Luna \citep{openai2026gpt56sol}. We also evaluate GPT-5.6 Luna as a source model handing off to Qwen 3.8 27B, testing whether handoff detection remains useful when the source is itself a lower-cost model. We evaluate both in-context and logfile transfer for each model pair. Table~\ref{tab:spilot-results} uses a fixed context-transfer method for each source model across both benchmarks and receiving models: logfile transfer for GPT-5.6 Sol and GPT-5.6 Luna, and in-context transfer for all other source models. Complete results for both transfer methods are provided in Appendix~\ref{sec:spilot-full-results}.

The results show that \spilot{} detects handoff points at an average of 26.6--41.1\% of the source model's trajectory on DeepSWE, compared with 55.6--68.9\% on Terminal-Bench, indicating that detected handoff points occur at different stages across benchmarks. \spilot{} substantially reduces API cost, with accuracy preservation varying across model pairs and benchmarks: 12 of the 22 (model-pair, benchmark) settings either improve accuracy or incur a relative loss of less than 5\%. For example, on DeepSWE, Kimi K3 handing off to Qwen 3.8 27B achieves 72.6\% accuracy compared with 73.5\% for capable-only execution, while reducing cost by 72.9\%. A stronger efficient model can further improve the outcome: with GPT-5.6 Luna, four of the five capable models exceed their capable-only accuracy on DeepSWE, while GPT-5.6 Sol incurs only a 1.2\% relative loss. These configurations reduce cost by 66.6--73.9\%; for instance, Opus-to-Luna improves accuracy from 67.3\% to 76.1\% while reducing cost by 70.5\%.
Even when the source model is already efficient, Luna-to-Qwen reduces API cost by 75.0\% on DeepSWE and 42.9\% on Terminal-Bench, with relative accuracy losses of 7.3\% and 5.7\% respectively, indicating the versatility of \spilot{}.

\subsection{Comparison with Routing and Handoff Baselines}
\label{sec:routing-baselines}
To put these results in context, we compare \spilot{} with query-level and per-turn routing as alternative ways.
We first test query-level routing by fine-tuning Qwen 3.5 2B to predict whether Qwen 3.8 27B can solve a task from its initial description, using success labels from 2,000 SWE-rebench V2 rollouts with balanced positive and negative examples. Despite 99.55\% acc. on the training set, the router selects Qwen for only one out of all tasks, thus it effectively reduces to capable-only execution and provides little cost savings.

We then reproduce MTRouter \citep{zhang2026mtrouter} with the same model pairs to investigate whether a more fine-grained, model-specific per-turn routing strategy can achieve better accuracy--cost trade-offs. As shown in Table~\ref{tab:routing-baselines}, \spilot{} reduces API cost by 46.7--86.6\% relative to MTRouter, while matching or improving accuracy in all four settings. 
These results indicate that our learned one-time handoff strategy achieves even better accuracy--cost trade-offs than per-turn routing in the evaluated settings.

\begin{wraptable}{r}{0.56\textwidth}
\vspace{-\intextsep}
\centering
\caption{\spilot{} versus routing and fixed handoff baselines with Qwen 3.8 27B. Fixed thresholds use the mean progress or turn under \spilot{} for each setting. Accuracy is in percent; cost is mean API cost per task.
}
\label{tab:routing-baselines}
\footnotesize
\setlength{\tabcolsep}{3pt}
\renewcommand{\arraystretch}{1.0}
\begin{tabular*}{\linewidth}{@{\extracolsep{\fill}}lrrrr@{}}
\toprule
 & \multicolumn{2}{c}{DeepSWE v1.1} & \multicolumn{2}{c}{Terminal Bench 2.1} \\
\cmidrule(lr){2-3}\cmidrule(lr){4-5}
Method & Acc. (\%) & Cost (\$) & Acc. (\%) & Cost (\$) \\
\midrule
\multicolumn{5}{c}{\textit{Capable: GLM 5.3; Efficient: Qwen 3.8 27B}} \\
MTRouter & 58.4 & 3.67 & 80.9 & 1.35 \\
Fixed progress & 59.3 & 0.52 & 78.7 & 0.38 \\
Fixed turn & 54.9 & 0.48 & 76.4 & 0.18 \\
\spilot{} & 58.4 & 0.49 & 80.9 & 0.27 \\
\midrule
\multicolumn{5}{c}{\textit{Capable: Kimi K3; Efficient: Qwen 3.8 27B}} \\
MTRouter & 66.4 & 4.37 & 85.4 & 1.07 \\
Fixed progress & 74.3 & 1.28 & 86.5 & 0.83 \\
Fixed turn & 69.9 & 1.15 & 82.0 & 0.37 \\
\spilot{} & 72.6 & 1.28 & 86.5 & 0.57 \\
\bottomrule
\end{tabular*}
\vspace{-10pt}
\end{wraptable}

We also compare \spilot{} with fixed-progress and fixed-turn handoffs, using its mean handoff progress and mean handoff turn for each benchmark and model pair as the respective thresholds. Fixed-progress handoff has access to future information, i.e., the source model's final trajectory length, that is unavailable during online execution. Table~\ref{tab:routing-baselines} reveals that despite this advantage, its accuracy remains close to \spilot{}, while \spilot{} incurs equal or lower API cost. Compared with fixed-turn handoff, \spilot{} achieves consistently higher accuracy. These results show that online detection can achieve accuracy comparable to a fixed-progress baseline with access to future information, while preserving accuracy better than a fixed turn threshold.

\subsection{Pushing the Accuracy--Cost Pareto Frontier}
\label{sec:pareto-frontier}

Finally, we examine whether \spilot{} can improve the accuracy--cost trade-offs obtained by varying a model's reasoning effort. On DeepSWE v1.1, we sweep \texttt{low}, \texttt{medium}, \texttt{high}, \texttt{xhigh}, and \texttt{max} for GPT-5.6 Sol and Claude Opus 5; \texttt{low} and \texttt{max} for GLM-5.3; \texttt{low}, \texttt{high}, and \texttt{max} for Kimi K3; and \texttt{high}, \texttt{xhigh}, and \texttt{max} for GPT-5.6 Luna. For each source configuration, we use \spilot{} to detect a handoff point and continue with Qwen 3.8 27B. We additionally evaluate continuations with GPT-5.6 Luna for the first four source models. We use logfile transfer for handoffs from GPT-5.6 Sol and GPT-5.6 Luna, and in-context transfer for all other source models. Figure~\ref{fig:reasoning-effort-tradeoff} compares these configurations with single-model baselines, and Appendix~\ref{sec:reasoning-effort-results} provides the complete numerical results.
The results show that \spilot{} extends the empirical accuracy--cost frontier beyond the evaluated single-model configurations across a range of budgets. At the low-cost end, Luna-to-Qwen with \texttt{xhigh} effort matches Luna's standalone accuracy of 56.6\% while reducing mean API cost from \$0.21 to approximately \$0.06 per task. At a higher budget, GLM-to-Luna with \texttt{max} effort achieves 72.6\% accuracy at \$0.80 per task, compared with 69.0\% at \$3.06 for GLM alone. Kimi-to-Luna with \texttt{high} effort reaches 79.6\% accuracy at \$1.60 per task, exceeding the highest single-model accuracy in the sweep, 75.2\% from Opus with \texttt{xhigh} effort at \$8.90. Different source models, effort settings, and receiving models contribute to the frontier, suggesting that reasoning effort and model handoff offer complementary ways to allocate computation over a task.

\section{Related work}
\label{sec:related}


\paragraph{Request-level routing and cascades.}
Prior work optimizes inference allocation at the request level.
For instance, FrugalGPT \citep{chen2023frugalgpt} and \citet{dekoninck2025cascade} explore cost-aware cascading pipelines, while RouteLLM \citep{ong2024routellm}, RouterDC \citep{chen2024routerdc}, and Route to Reason \citep{pan2026routereason} learn query-level model selection, compatibility, or reasoning strategies, supported by unified evaluation benchmarks \citep{hu2024routerbench}.
These approaches primarily address individual requests. In interactive agentic tasks, an agent's actions modify the workspace and accumulate interaction history, changing the state inherited by a subsequent model.
Consequently, determining which model can resolve this evolving state requires accounting for task progress beyond the original request.

\paragraph{Multi-round routing and orchestration.}
Beyond single-turn decisions, multi-round orchestration architectures dynamically coordinate model and tool execution.
Frameworks such as Router-R1 \citep{zhang2025routerr1}, ToolOrchestra \citep{su2025toolorchestra}, and SkillOrchestra \citep{wang2026skillorchestra} learn competence, invocation patterns, or skill-specific costs from execution feedback.
Similarly, MTRouter \citep{zhang2026mtrouter} uses interaction history for turn-level routing, while Agent-as-a-Router \citep{zhou2026agentrouter} draws on verified experience across a stream of coding tasks.
These approaches select among candidate models or tools, sometimes using model descriptions to support generalization to unseen candidates, as demonstrated by Router-R1 and ToolOrchestra.
\spilot{} separates handoff timing from model selection: for a given capable--efficient model pair, a lightweight learned detector assesses task progress after each turn and decides whether to transfer control. We evaluate the same detector across multiple model pairs, with the receiving model continuing until termination after a handoff.

\paragraph{Model handoffs and trajectory transfer.}
A closely related line of research uses execution feedback to inform model transitions.
SWE-Router \citep{son2026swerouter} uses an exploratory trajectory to inform escalation, while TACIT-Switch \citep{lei2026tacit} learns permanent small-to-large handoffs based on accumulated risk signals.
The Handoff Tax \citep{ganz2026handofftax} studies how handoff direction, timing, and trajectory-transfer interfaces affect coding-agent performance, finding direction-dependent accuracy--cost trade-offs. Its experiments use predefined switch points to isolate the effects of handoff and inherited history.
Our analysis likewise examines handoff direction and context transfer, and further motivates a learned policy that identifies suitable capable-to-efficient handoff points online based on recurring patterns observed in agent trajectories.

\paragraph{Switching within a reasoning output.}
Finally, several approaches perform fine-grained handoffs within a single reasoning trajectory.
STEER \citep{lee2025steer} and SplitReason \citep{akhauri2025splitreason} offload reasoning steps based on confidence or learned difficulty, while Speculative Thinking \citep{yang2025specthinking} delegates reflective reasoning to larger models.
Most relevantly, RelayGen \citep{song2026relaygen} transfers simpler generation segments from a strong model to a smaller one via model-specific cues without router training.
Our handoffs occur between complete agent turns, involving tool use, a shared workspace, and persistent changes to the task state.
The receiving model inherits this workspace and accesses prior interaction history through in-context or logfile transfer, then continues acting until termination. Our focus is on how task progress and context transfer support continuation across models.



\section{Conclusion}
We study how capable and efficient models can collaborate within a single long-horizon agentic task through a one-time handoff. Our experiments show that capable-to-efficient handoffs offer favorable accuracy--cost trade-offs, particularly once the capable model has established an implementation, gathered feedback, and narrowed the remaining work to localized refinement. Motivated by these findings, we introduce \spilot{}, a lightweight detector that dynamically identifies suitable handoff points. Across DeepSWE v1.1 and Terminal-Bench 2.1, \spilot{} substantially reduces API cost while largely preserving, and occasionally improving, task accuracy. Further reasoning-effort sweeps demonstrate that such handoffs can extend the empirical Pareto frontier beyond single-model configurations, highlighting task progress as an effective signal for dynamic capability allocation.

\newpage
\label{main-text-end}

\bibliography{ref}
\bibliographystyle{unsrtnat}
\clearpage
\appendix
\section{Handoff progress on Terminal-Bench}
\label{sec:terminalbench-handoff}
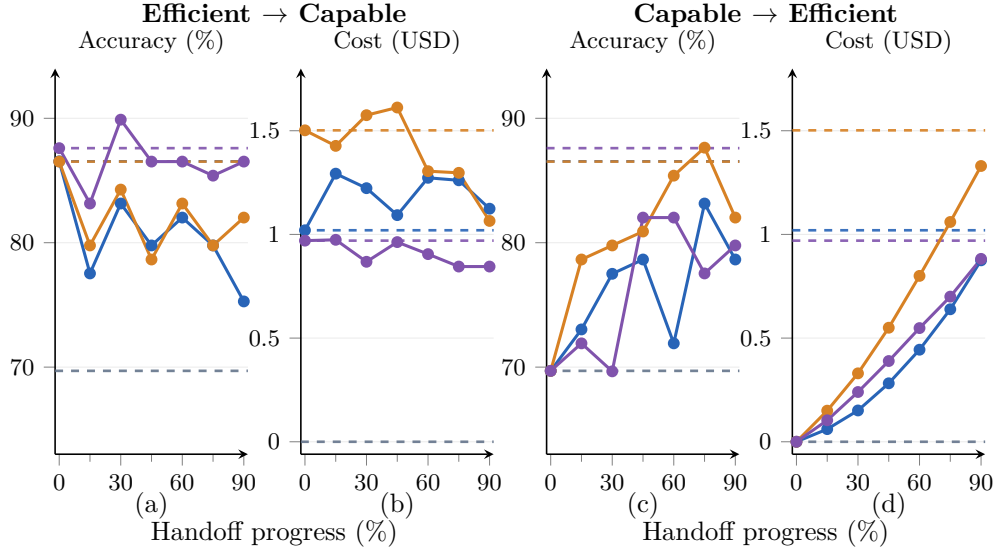
\begin{figure}[t]
\centering
\begin{tikzpicture}
\definecolor{glm}{HTML}{2864B7}
\definecolor{kimi}{HTML}{D78124}
\definecolor{gpt}{HTML}{8053AC}
\definecolor{qwen}{HTML}{64748B}
\begin{groupplot}[group style={group size=4 by 1,horizontal sep=.7cm},width=2.55cm,height=5.1cm,scale only axis,xmin=-2,xmax=92,xtick={0,30,60,90},minor xtick={15,45,75},tick label style={font=\fontsize{9}{10}\selectfont},label style={font=\fontsize{9}{10}\selectfont},title style={font=\fontsize{9}{10}\selectfont,yshift=-4pt},axis line style={line width=.6pt},axis lines=left,ymajorgrids,grid style={gray!15},tick align=outside,scaled ticks=false]
\nextgroupplot[ymin=63,ymax=94,title={Accuracy (\%)}]
\addplot[glm,dashed,line width=1.0pt,opacity=.9,forget plot] coordinates {(-2,86.54) (92,86.54)};
\addplot[kimi,dashed,line width=1.0pt,opacity=.9,forget plot] coordinates {(-2,86.52) (92,86.52)};
\addplot[gpt,dashed,line width=1.0pt,opacity=.9,forget plot] coordinates {(-2,87.6) (92,87.6)};
\addplot[qwen,dashed,line width=1.0pt,opacity=.9,forget plot] coordinates {(-2,69.7) (92,69.7)};
\addplot[glm,line width=1.15pt,mark=*,mark size=1.65pt] coordinates {(0,86.54) (15,77.53) (30,83.15) (45,79.78) (60,82.02) (75,79.78) (90,75.28)};
\addplot[kimi,line width=1.15pt,mark=*,mark size=1.65pt] coordinates {(0,86.52) (15,79.78) (30,84.27) (45,78.65) (60,83.15) (75,79.78) (90,82.02)};
\addplot[gpt,line width=1.15pt,mark=*,mark size=1.65pt] coordinates {(0,87.6) (15,83.15) (30,89.89) (45,86.52) (60,86.52) (75,85.39) (90,86.52)};
\nextgroupplot[ymin=-0.06,ymax=1.8,title={Cost (USD)},yticklabel style={/pgf/number format/fixed,/pgf/number format/precision=1}]
\addplot[glm,dashed,line width=1.0pt,opacity=.9,forget plot] coordinates {(-2,1.02) (92,1.02)};
\addplot[kimi,dashed,line width=1.0pt,opacity=.9,forget plot] coordinates {(-2,1.502) (92,1.502)};
\addplot[gpt,dashed,line width=1.0pt,opacity=.9,forget plot] coordinates {(-2,0.97) (92,0.97)};
\addplot[qwen,dashed,line width=1.0pt,opacity=.9,forget plot] coordinates {(-2,0.0) (92,0.0)};
\addplot[glm,line width=1.15pt,mark=*,mark size=1.65pt] coordinates {(0,1.02) (15,1.293) (30,1.223) (45,1.093) (60,1.273) (75,1.261) (90,1.124)};
\addplot[kimi,line width=1.15pt,mark=*,mark size=1.65pt] coordinates {(0,1.502) (15,1.427) (30,1.575) (45,1.612) (60,1.305) (75,1.297) (90,1.065)};
\addplot[gpt,line width=1.15pt,mark=*,mark size=1.65pt] coordinates {(0,0.97) (15,0.974) (30,0.868) (45,0.963) (60,0.905) (75,0.845) (90,0.845)};
\nextgroupplot[ymin=63,ymax=94,title={Accuracy (\%)}]
\addplot[glm,dashed,line width=1.0pt,opacity=.9,forget plot] coordinates {(-2,86.54) (92,86.54)};
\addplot[kimi,dashed,line width=1.0pt,opacity=.9,forget plot] coordinates {(-2,86.52) (92,86.52)};
\addplot[gpt,dashed,line width=1.0pt,opacity=.9,forget plot] coordinates {(-2,87.6) (92,87.6)};
\addplot[qwen,dashed,line width=1.0pt,opacity=.9,forget plot] coordinates {(-2,69.7) (92,69.7)};
\addplot[glm,line width=1.15pt,mark=*,mark size=1.65pt] coordinates {(0,69.7) (15,73.03) (30,77.5) (45,78.65) (60,71.91) (75,83.15) (90,78.65)};
\addplot[kimi,line width=1.15pt,mark=*,mark size=1.65pt] coordinates {(0,69.7) (15,78.65) (30,79.78) (45,80.9) (60,85.39) (75,87.64) (90,82.02)};
\addplot[gpt,line width=1.15pt,mark=*,mark size=1.65pt] coordinates {(0,69.7) (15,71.91) (30,69.66) (45,82.02) (60,82.02) (75,77.53) (90,79.78)};
\nextgroupplot[ymin=-0.06,ymax=1.8,title={Cost (USD)},yticklabel style={/pgf/number format/fixed,/pgf/number format/precision=1}]
\addplot[glm,dashed,line width=1.0pt,opacity=.9,forget plot] coordinates {(-2,1.02) (92,1.02)};
\addplot[kimi,dashed,line width=1.0pt,opacity=.9,forget plot] coordinates {(-2,1.502) (92,1.502)};
\addplot[gpt,dashed,line width=1.0pt,opacity=.9,forget plot] coordinates {(-2,0.97) (92,0.97)};
\addplot[qwen,dashed,line width=1.0pt,opacity=.9,forget plot] coordinates {(-2,0.0) (92,0.0)};
\addplot[glm,line width=1.15pt,mark=*,mark size=1.65pt] coordinates {(0,0.0) (15,0.061) (30,0.151) (45,0.282) (60,0.444) (75,0.639) (90,0.875)};
\addplot[kimi,line width=1.15pt,mark=*,mark size=1.65pt] coordinates {(0,0.0) (15,0.15) (30,0.33) (45,0.55) (60,0.8) (75,1.06) (90,1.33)};
\addplot[gpt,line width=1.15pt,mark=*,mark size=1.65pt] coordinates {(0,0.0) (15,0.1036) (30,0.24) (45,0.3894) (60,0.5483) (75,0.7) (90,0.8821)};
\end{groupplot}
\node[font=\fontsize{10}{11}\selectfont] at ([yshift=-.68cm]group c1r1.south) {(a)};
\node[font=\fontsize{10}{11}\selectfont] at ([yshift=-.68cm]group c2r1.south) {(b)};
\node[font=\fontsize{10}{11}\selectfont] at ([yshift=-.68cm]group c3r1.south) {(c)};
\node[font=\fontsize{10}{11}\selectfont] at ([yshift=-.68cm]group c4r1.south) {(d)};
\path (group c1r1.north west) -- (group c2r1.north east) coordinate[midway] (row1);
\node[font=\fontsize{10}{11}\selectfont\bfseries] at ([yshift=.7cm]row1) {Efficient $\to$ Capable};
\path (group c1r1.south west) -- (group c2r1.south east) coordinate[midway] (xlabel1);
\node[font=\fontsize{10}{11}\selectfont] at ([yshift=-1.05cm]xlabel1) {Handoff progress (\%)};
\path (group c3r1.north west) -- (group c4r1.north east) coordinate[midway] (row2);
\node[font=\fontsize{10}{11}\selectfont\bfseries] at ([yshift=.7cm]row2) {Capable $\to$ Efficient};
\path (group c3r1.south west) -- (group c4r1.south east) coordinate[midway] (xlabel2);
\node[font=\fontsize{10}{11}\selectfont] at ([yshift=-1.05cm]xlabel2) {Handoff progress (\%)};
\end{tikzpicture}
\caption[Handoff progress on Terminal-Bench 2.1]{Accuracy and mean API cost across handoff progress on Terminal-Bench 2.1. Colors identify the capable model: \handoffline{handoffGLM}{solid} GLM-5.3, \handoffline{handoffKimi}{solid} Kimi K3, and \handoffline{handoffGPT}{solid} GPT-5.6 Sol; the efficient model is Qwen 3.8 27B. Solid curves show handoffs using logfile transfer in (a--b); in (c--d), GLM-5.3 and Kimi K3 use in-context transfer, while GPT-5.6 Sol uses logfile transfer. Horizontal dashed lines (\handoffline{handoffGLM}{dashed}\,\handoffline{handoffKimi}{dashed}\,\handoffline{handoffGPT}{dashed}) show the corresponding capable-only baselines; \handoffline{handoffQwen}{dashed} denotes Qwen-only execution. Progress is relative to the source model's completed baseline trajectory; 0\% denotes receiving-model-only execution.}
\label{fig:handoff-progress-tb}
\setcounter{handoffpanel}{0}
\refstepcounter{handoffpanel}\label{fig:handoff-e2c-tb-accuracy}
\refstepcounter{handoffpanel}\label{fig:handoff-e2c-tb-cost}
\refstepcounter{handoffpanel}\label{fig:handoff-c2e-tb-accuracy}
\refstepcounter{handoffpanel}\label{fig:handoff-c2e-tb-cost}
\end{figure}

Figure~\ref{fig:handoff-progress-tb} reports the Terminal-Bench 2.1 results using the same handoff checkpoints as the DeepSWE experiments in the main text. As in the main text, we use logfile transfer for GPT-5.6 Sol handing off to Qwen 3.8 27B. For efficient-to-capable handoffs, accuracy and cost fluctuate across checkpoints, and Qwen-to-GLM handoffs cost more than GLM alone throughout the sweep. Capable-to-efficient handoffs offer favorable trade-offs at some checkpoints: Kimi K3 handing off at 75\% reaches 87.64\% accuracy versus 86.52\% for Kimi alone, with 29.4\% lower cost. However, the results are less consistent across models and checkpoints than on DeepSWE, and later handoffs do not monotonically improve accuracy.

\section{Handoff-point annotation protocol}
\label{sec:handoff-annotation}
We use Codex to annotate the earliest suitable handoff point in each collected trajectory. The protocol follows the qualitative patterns in Section~\ref{sec:case_study}. Codex reviews each trajectory retrospectively to better locate the earliest suitable handoff point.

\paragraph{Evidence available to the annotator.}
The annotator reads the task description and the chronological sequence of tool calls and their responses. Private reasoning text, post-trajectory verifier outputs, final rewards, and outcomes of receiving-model continuations are excluded as annotation evidence.

\paragraph{Readiness criteria.}
The annotator selects the earliest turn at which the evidence establishes all three conditions:
\begin{enumerate}
    \item \textbf{A substantive artifact exists.} There is an implementation, runnable prototype, or task-required artifact. Investigation, a plan, or an unexecuted code sketch alone is insufficient.
    \item \textbf{The artifact has received real feedback.} It has been compiled, executed, tested, or inspected by a task-relevant checker, with an observed response. Issuing a command does not establish that it succeeded. A failed check can qualify when it exposes a localized problem; an environment failure that prevents the artifact from being exercised does not by itself establish this condition.
    \item \textbf{The remaining work has become localized.} The evidence indicates that the work ahead mainly concerns diagnosis, repair, hardening, validation, or final-output preparation, rather than selecting the core approach or implementing the main solution from scratch.
\end{enumerate}
These conditions do not require all tests to pass. Conversely, a passing check on an unrelated component or a substitute that bypasses the central requested behavior is insufficient. The annotator considers what the observed check actually establishes about the task-required artifact.

\paragraph{Selecting the first point and handling uncertainty.}
The label records the first qualifying point, not the final successful test or submission. If later work introduces a regression or substantially rewrites the implementation, the earlier point is retained and the instability is noted. If the conditions are never jointly established, the trajectory is labeled \texttt{no\_clear\_handoff\_point}, with no handoff index. Missing or unreadable trajectories are recorded separately and are not assigned this label. Readable failed or interrupted trajectories remain eligible for annotation. The provenance of an implementation does not by itself determine readiness; when relevant, it is documented as a caveat.

\paragraph{Turn indexing and annotation record.}
A turn consists of one assistant step containing one or more tool calls together with their responses. Both the handoff index and evidence indices are zero-based within these tool-bearing steps, rather than indices over all conversation messages. We also preserve each step's original identifier to support traceability to the source trajectory. Each record includes the task and source model, the actual trajectory path, the total number of tool-bearing turns, the readiness status, the handoff index and source step identifier, supporting tool evidence, a brief rationale, a confidence level, and any caveats. For trajectories without a clear handoff point, the handoff fields are null.

\section{Complete \spilot{} results}
\label{sec:spilot-full-results}
Table~\ref{tab:spilot-full-results} reports both in-context and logfile transfer for all five capable models and both efficient models, as well as GPT-5.6 Luna handing off to Qwen 3.8 27B. The main-text table uses logfile transfer for GPT-5.6 Sol and GPT-5.6 Luna as source models, and in-context transfer for all other source models. Single-model baselines are reported in Table~\ref{tab:spilot-results}.
\begin{table}[t]
\centering
\caption{Complete \spilot{} results for both context-transfer methods. Each row reports accuracy and mean API cost per task from the same configuration.}
\label{tab:spilot-full-results}
\small
\setlength{\tabcolsep}{2pt}
\renewcommand{\arraystretch}{1.0}
\begin{tabular}{@{}lllrrrr@{}}
\toprule
 & & & \multicolumn{2}{c}{DeepSWE v1.1} & \multicolumn{2}{c}{Terminal-Bench 2.1}\\
\cmidrule(lr){4-5}\cmidrule(l){6-7}
Capable model & Efficient model & Transfer & Acc. (\%) & Cost (\$) & Acc. (\%) & Cost (\$)\\
\midrule
GPT-5.6 Sol & Qwen 3.8 27B & In-context & 54.9 & 1.03 & 85.4 & 0.43 \\
 &  & Logfile & 60.2 & 1.02 & 82.0 & 0.43 \\
 & GPT-5.6 Luna & In-context & 61.9 & 1.27 & 84.3 & 0.47 \\
 &  & Logfile & 67.3 & 1.39 & 82.0 & 0.47 \\
\midrule
Claude Opus 5 & Qwen 3.8 27B & In-context & 61.9 & 2.61 & 87.6 & 1.13 \\
 &  & Logfile & 61.1 & 2.44 & 87.6 & 1.06 \\
 & GPT-5.6 Luna & In-context & 76.1 & 2.92 & 86.5 & 1.17 \\
 &  & Logfile & 70.8 & 2.99 & 85.4 & 1.19 \\
\midrule
GLM-5.3 & Qwen 3.8 27B & In-context & 58.4 & 0.49 & 80.9 & 0.27 \\
 &  & Logfile & 53.1 & 0.49 & 78.7 & 0.26 \\
 & GPT-5.6 Luna & In-context & 66.4 & 0.83 & 79.8 & 0.33 \\
 &  & Logfile & 67.3 & 0.87 & 85.4 & 0.34 \\
\midrule
Kimi K3 & Qwen 3.8 27B & In-context & 72.6 & 1.28 & 86.5 & 0.57 \\
 &  & Logfile & 65.5 & 1.19 & 84.3 & 0.53 \\
 & GPT-5.6 Luna & In-context & 77.0 & 1.58 & 80.9 & 0.62 \\
 &  & Logfile & 74.3 & 1.64 & 85.4 & 0.61 \\
\midrule
Gemini 3.8 Flash & Qwen 3.8 27B & In-context & 61.9 & 1.21 & 86.5 & 0.51 \\
 &  & Logfile & 54.0 & 1.20 & 80.9 & 0.51 \\
 & GPT-5.6 Luna & In-context & 73.5 & 1.71 & 85.4 & 0.56 \\
 &  & Logfile & 68.1 & 1.64 & 85.4 & 0.56 \\
\midrule
GPT-5.6 Luna & Qwen 3.8 27B & In-context & 54.9 & 0.10 & 74.2 & 0.04 \\
 & & Logfile & 55.8 & 0.10 & 77.5 & 0.04 \\
\bottomrule
\end{tabular}
\end{table}

\section{Complete reasoning-effort sweep results}
\label{sec:reasoning-effort-results}
Table~\ref{tab:reasoning-effort-results} lists all single-model and \spilot{} configurations plotted in Figure~\ref{fig:reasoning-effort-tradeoff}, including dominated configurations. Qwen and Luna denote Qwen 3.8 27B and GPT-5.6 Luna, respectively. The Qwen baseline uses a 1M-token context window.
\begin{table}[t]
\centering
\caption{Complete numerical results for the reasoning-effort sweep on DeepSWE v1.1 in Figure~\ref{fig:reasoning-effort-tradeoff}. Effort refers to the source model. Accuracy is in percent and cost is mean API cost in USD per task. Dashes indicate configurations not evaluated in this sweep.}
\label{tab:reasoning-effort-results}
\footnotesize
\setlength{\tabcolsep}{4pt}
\renewcommand{\arraystretch}{1.05}
\begin{tabular}{llrrrrrr}
\toprule
 & & \multicolumn{2}{c}{Single model} & \multicolumn{2}{c}{\spilot{} $\to$ Qwen} & \multicolumn{2}{c}{\spilot{} $\to$ Luna} \\
\cmidrule(lr){3-4}\cmidrule(lr){5-6}\cmidrule(lr){7-8}
Source model & Effort & Acc. & Cost & Acc. & Cost & Acc. & Cost \\
\midrule
GPT-5.6 Sol & \texttt{low} & 35.4 & 0.65 & 47.8 & 0.20 & 65.5 & 0.60 \\
 & \texttt{medium} & 51.3 & 1.26 & 54.9 & 0.30 & 68.1 & 0.80 \\
 & \texttt{high} & 62.8 & 1.88 & 57.5 & 0.50 & 62.8 & 0.90 \\
 & \texttt{xhigh} & 71.7 & 2.79 & 59.3 & 0.70 & 69.0 & 1.10 \\
 & \texttt{max} & 71.7 & 4.84 & 61.9 & 1.00 & 70.8 & 1.40 \\
\midrule
Claude Opus 5 & \texttt{low} & 54.0 & 1.45 & 49.6 & 0.70 & 65.5 & 1.00 \\
 & \texttt{medium} & 68.1 & 3.14 & 55.8 & 1.10 & 75.2 & 1.40 \\
 & \texttt{high} & 64.6 & 5.87 & 59.3 & 1.70 & 69.9 & 2.00 \\
 & \texttt{xhigh} & 75.2 & 8.90 & 55.8 & 2.20 & 70.8 & 2.60 \\
 & \texttt{max} & 67.3 & 9.91 & 61.9 & 2.61 & 76.1 & 2.92 \\
\midrule
GLM-5.3 & \texttt{low} & 59.3 & 1.71 & 61.9 & 0.29 & 67.3 & 0.60 \\
 & \texttt{max} & 69.0 & 3.06 & 66.4 & 0.48 & 72.6 & 0.80 \\
\midrule
Kimi K3 & \texttt{low} & 54.9 & 1.40 & 54.0 & 0.55 & 72.6 & 0.90 \\
 & \texttt{high} & 69.9 & 3.81 & 71.7 & 1.20 & 79.6 & 1.60 \\
 & \texttt{max} & 73.5 & 4.73 & 72.6 & 1.28 & 77.0 & 1.58 \\
\midrule
Gemini 3.8 Flash & \texttt{high} & 65.5 & 6.15 & --- & --- & --- & --- \\
\midrule
Qwen 3.8 27B & \texttt{xhigh} & 29.2 & 0.00 & --- & --- & --- & --- \\
\midrule
GPT-5.6 Luna & \texttt{high} & 38.1 & 0.11 & 50.4 & 0.04 & --- & --- \\
 & \texttt{xhigh} & 56.6 & 0.21 & 56.6 & 0.06 & --- & --- \\
 & \texttt{max} & 60.2 & 0.40 & 55.8 & 0.10 & --- & --- \\
\bottomrule
\end{tabular}
\end{table}

\end{document}